\documentclass[11pt,twoside]{article}
\usepackage{CAGN2019}
\usepackage{graphicx}

\usepackage{latexsym}
\usepackage{verbatim}

\usepackage{ifpdf}  
\ifpdf  
      \DeclareGraphicsExtensions{.pdf,.png,.jpg}  
\else  
      \DeclareGraphicsExtensions{.eps}  
\fi 

\begin{document}

\vskip 1.0cm
\markboth{W.~Maciel \& S. Andrievsky}{Galactic radial abundance gradients}
\pagestyle{myheadings}
%
%
\vspace*{0.5cm}
\parindent 0pt{Invited Review}


\vspace*{0.5cm}
\title{Galactic radial abundance gradients: cepheids and photoionized nebulae}

\author{W.J.~Maciel$^1$, S.~Andrievsky$^2$}
\affil{$^1$Astronomy Department, University of S\~ao Paulo, S\~ao Paulo, Brazil\\
$^2$Odessa National University, Ukraine}

\begin{abstract}

Radial abundance gradients are observed in the Galaxy and other galaxies as well, and include 
several chemical elements in different stellar systems. Possibly the most accurate gradients 
in the Galaxy are those determined from chepheid variable stars. These objects have very accurate 
abundances for many elements and are generally considered as standard candles, so that their 
galactocentric distances are very well determined. These stars are relatively young, with ages 
between the main types of photoionized nebulae, namely the younger HII regions and the older 
planetary nebulae. In this paper we consider the O/H and Fe/H gradients based on a large sample 
of galactic cepheids, and compare the results with recent determinations from photoionized nebulae.


\end{abstract}

\section{Introduction}

Radial abundance gradients are increasingly important as a constraint of chemical evolution models 
both for the Galaxy and other galaxies as well. They can be measured using a variety of objects, 
such as HII regions, cepheid variables, planetary nebulae, open clusters, etc. Since these objects 
comprise a large range of ages, from a few million years to several Gyr, the gradients can also 
give some information on the time variation of the abundances in several galactic systems.

It is generally believed that the gradients derived from young objects, such as HII regions and 
cepheid stars are somewhat different from those measured in older systems, such as planetary nebulae 
and open clusters. However, the interpretation of the data is complex, in view of the different 
elements that can be studied and of the considerable uncertainties in assigning a definite age to 
each object, especially in the case of the older systems. Some references include \cite{anders}, 
\cite{molla2018, molla2019}, \cite{mc2014, mc2013}, \cite{mrc2012}, \cite{cmcm2014, ccm2011}. Some 
recent results obtained by our group suggest that the gradients have not appreciably changed in 
the past 3-5 Gyr, although these results are dependent on relatively uncertain age estimates 
[\cite{mrc2012}, \cite{mc2013}].

Another source of uncertainties are possible space variations of the gradients, such as the proposed 
flattening at large galactocentric distances, the behaviour of the gradients in the inner Galaxy,
which can be affected by the presence of bars, and near the solar galactocentric distance, where a 
large amount of data is available. The simplest models assume a unique linear gradient throughout 
the galactic disk, but more complicated variations have also been considered, such as multiple linear 
gradients in different parts of the disk and non linear variations [see for example \cite{mcc2015}, 
\cite{davies}, \cite{ivezic},  \cite{esteban2017}].

In this paper, we consider the O/H and Fe/H radial gradients in the galactic disk as measured  by 
cepheid variable stars. There is a large amount of data for these stars, including the results of 
the group by S. Andrievsky and collaborators [\cite{serguei2016, serguei2014, serguei2013, serguei2004, 
serguei2002a, serguei2002b, serguei2002c}, \cite{luck2013, luck2011, luck2006, luck2003}, 
\cite{lucklam}, \cite{korotin}, \cite{martin}], as well as by other groups, [\cite{genovali2015a, 
genovali2015b, genovali2014, genovali2013}, \cite{lemasle2018, lemasle2015, lemasle2013}, 
\cite{pedicelli}]. Our goal is to consider the largest possible sample of reliable spectroscopic 
abundances as well as accurate galactocentric distances. It is expected that the use of a nearly 
complete sample may lead to an accurate determination of the gradients, so that the non-homogeneity 
of the data may be counterbalanced. We will analyze several possibilities  of the abundance 
variations in the disk, such as a unique linear gradient, multiple gradients, and non-linear 
variations. The results can then be compared with the results from other young and older systems, 
such as HII regions, planetary nebulae and open clusters.

\section{Cepheid gradients: O/H and Fe/H}

\subsection{The data}

The basic sample comes from the work by \cite{korotin}, who have presented a NLTE analysis of the 
oxygen abundance distribution in the thin disk based on infrared data from two telescopes: the 
Hobby-Eberly Telescope (HET) and the Max Planck Gesellschaft Telescope (MPG). The lines used are the 
triplet 777.1 - 777.4 nm. The HET data are described by \cite{lucklam}, whereas the MPG data are from 
\cite{luck2013} with atmospheric parameters from \cite{luck2011}. Fe/H data are also included from 
\cite{lucklam} and \cite{luck2013}. The lowest uncertainties in the O/H abundances are in the range 
$0.05$ to $0.08$ dex, and an average uncertainty is $0.12$. 

The data by \cite{lucklam} include Fe abundances with errors and oxygen abundances, but not distances. 
The average uncertainties for the Fe/H data are 0.141 dex for FeI and 0.119 dex for FeII, with a 
standard deviation of 0.047 dex, considering all. We adopt the FeII data when possible, so that we 
can assume and average uncertainty of 0.12 dex for the Fe/H abundances, which is the same as the O/H 
uncertainties in the data by \cite{korotin}. The distances and galactocentric distances of the objects 
in the sample by \cite{lucklam} are given in \cite{luck2013}. 

\cite{martin} presented chemical abundances for a sample of galactic cepheids, including also the 
elements O/H and Fe/H. The galactocentric distances range mostly from 5 to 7 kpc, adopting $R_0 = 
7.9\,$kpc for the solar galactocentric distance, that is, the objects are generally closer to the 
galactic centre than in the previous samples. A combination of the previous data with the present 
results suggests a plateau in the chemical abundances closer to the galactic centre.  The observations
were secured with the 3.6 CFHT telescope, the VLT, and the MPG telescope as in \cite{korotin}. Stellar 
atmosphere parameters are presented, as in the previous samples. Fe abundances are from LTE analysis, 
while for O/H NLTE is used. The uncertainties in the abundances are small, similar or lower than in 
\cite{korotin}. A value of 0.12 dex can be estimated both for O/H and Fe/H, being usually an upper limit. 

More recently, \cite{serguei2016} presented a detailed study of a cepheid variable star located closer 
to the galactic bulge than the previous samples. It is the object ASAS 181024-2049.6, for which they 
derive $R_G = 2.53$ kpc, $\epsilon$(O) = 9.17 and $\epsilon$(Fe) = 7.94, corresponding to [O/H] = 0.46 
and [Fe/H] = 0.44. The data were obtained with the CFHT telescope and a selection of 4 candidate stars 
was considered in the literature, one of which satisfied the usual criteria to distinguish Cepheid 
variables from W Vir stars. \cite{serguei2016} included this object in order to obtain a more complete 
sample than the one by \cite{martin}. This sample emphasized the objects closer to the galactic centre 
in order to investigate an apparent flattening in the gradients in this region, possibly as a 
consequence of the existence of a bar. Such a suggestion had already been made in the literature based 
on theoretical models and planetary nebula data [see for example \cite{cmcm2014, ccm2011}. In 
the paper, Andrievsky et al. consider a large number of elements such as Mg, Si, S, Ca, and Ti, all of 
which share the same characteristics.  

\cite{genovali2015a, genovali2015b} presented Fe/H abundances of a sample of galactic cepheids 
with data analyzed by \cite{genovali2014}. Some of the stars in their sample are also presented in the 
previous samples, so that they have not been considered. The data are based on high-resolution UVES 
spectra collected at ESO VLT (Cerro Paranal, Chile). The abundances are based essentially on the 
equivalent widths of FeI and FeII lines. The galactocentric distances are based on NIR photometry 
together with reddening-free Period-Wesenheit relations, with $R_0 = 7.94$ kpc. Some results by 
\cite{lemasle2018} have also been taken into account [see also \cite{lemasle2013, 
lemasle2015}]. This work includes double-mode cepheids from the Gaia Data Release (DR2). The 
metallicity is derived from the ratio of the first overtone and fundamental periods by Gaia DR2 and  
the parallaxes are used to determine the Galactocentric distances of the stars. The derived abundances 
are then used to investigate the effects on the galactic [Fe/H] gradient.

\subsection{Results}

We have analyzed a large sample of galactic cepheids with accurate abundances and distances. In this 
paper we report some results for a total sample of 361 independent stars with Fe/H 
abundances, and 331 stars with  O/H abundances. The main results are shown in Figures~1ab, 2ab, 
3ab, and 4ab, and in Tables 1 and 2. We adopt an average uncertainty of 0.12 dex both for O/H and 
Fe/H. From the discussion by \cite{groenewegen} the uncertainty in the galactocentric distances is 
very small, as cepheids are considered as standard candles. The average uncertainty in the distance 
for the 128 galactic cepheids of \cite{groenewegen} is about 5.38\%, and in the case of the 
galactocentric distance $R_G$ it is 0.79\%. Very few stars have uncertainties higher than 1\%, so that 
we adopted here an average uncertainty of 1\%. Figure~1a and 1b show the data with error bars for 
Fe/H and O/H, respectively. Figure~2ab includes linear fits, and the corresponding coefficients are 
given in Table~1, where we have
%
\begin{figure}
\begin{center}
\hspace{0.25cm}
\begin{tabular}{cc}
\includegraphics[height=7.0cm,angle=-90]{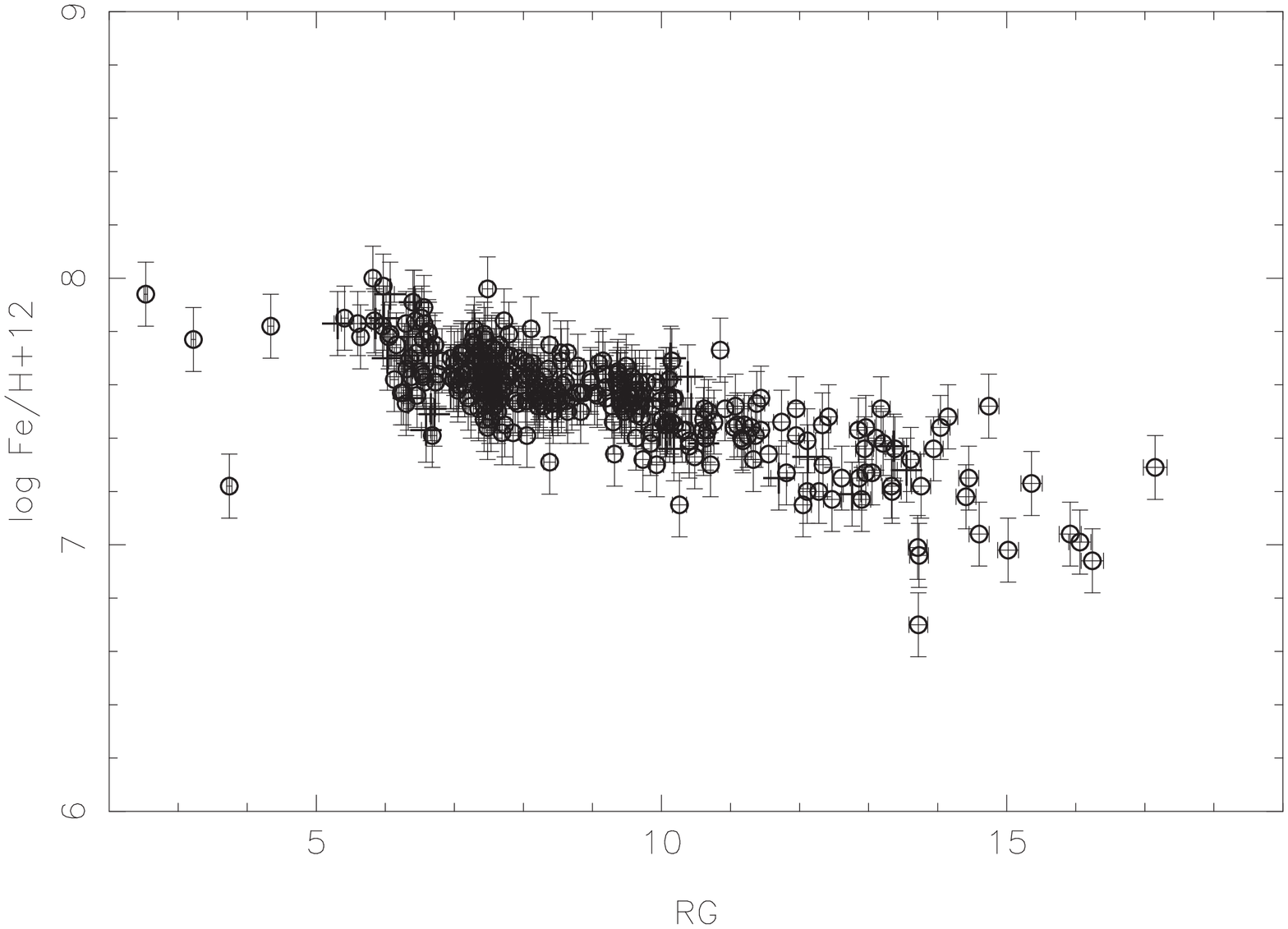} &
\includegraphics[height=7.0cm,angle=-90]{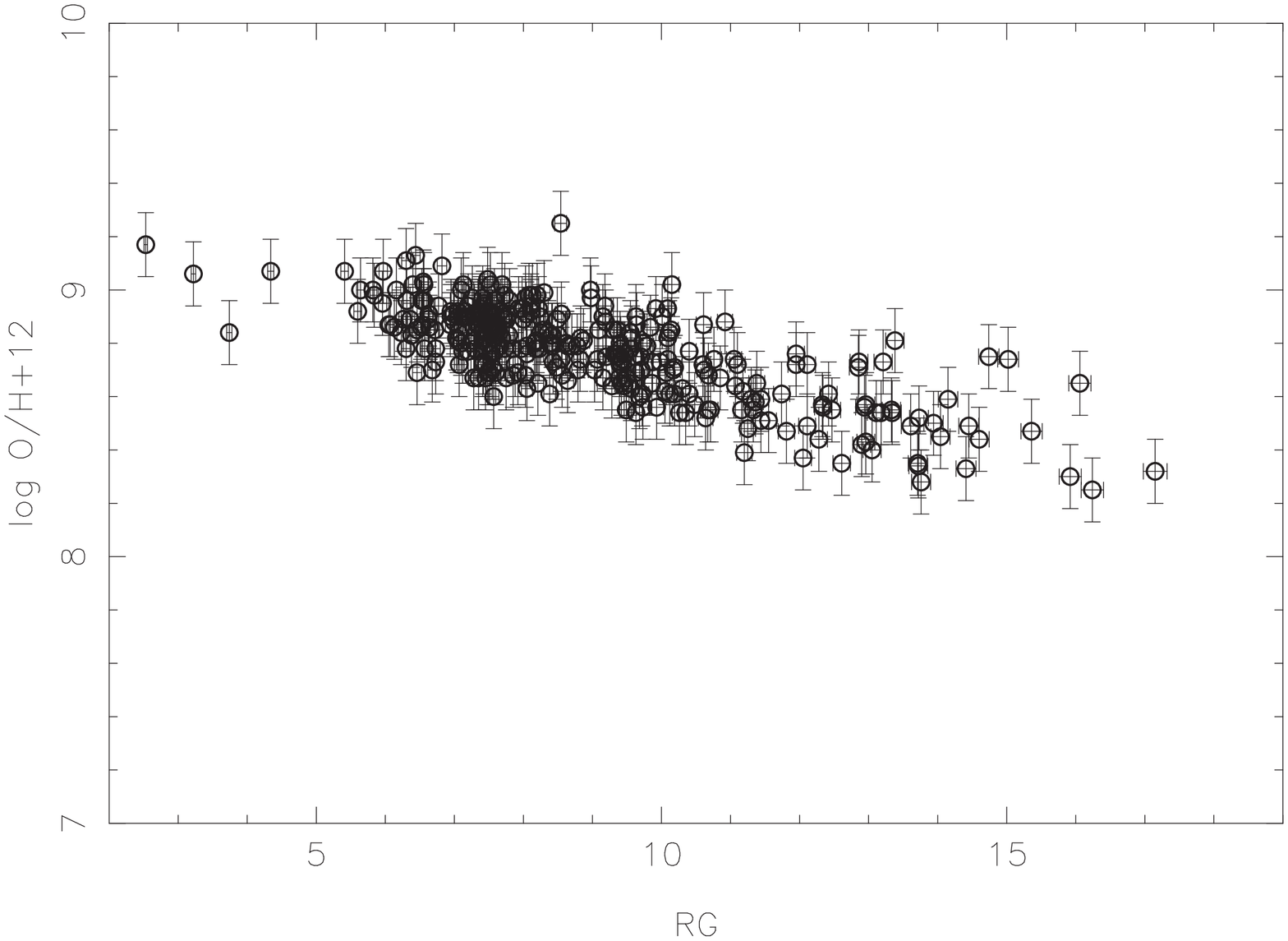}\\
\end{tabular}
\caption{Fe/H and O/H abundances functions of the galactocentric distance for galactic cepehids with 
the adopted error bars.}
\label{fig1ab}
\end{center}
%
\begin{center}
\hspace{0.25cm}
\begin{tabular}{cc}
\includegraphics[height=7.0cm,angle=-90]{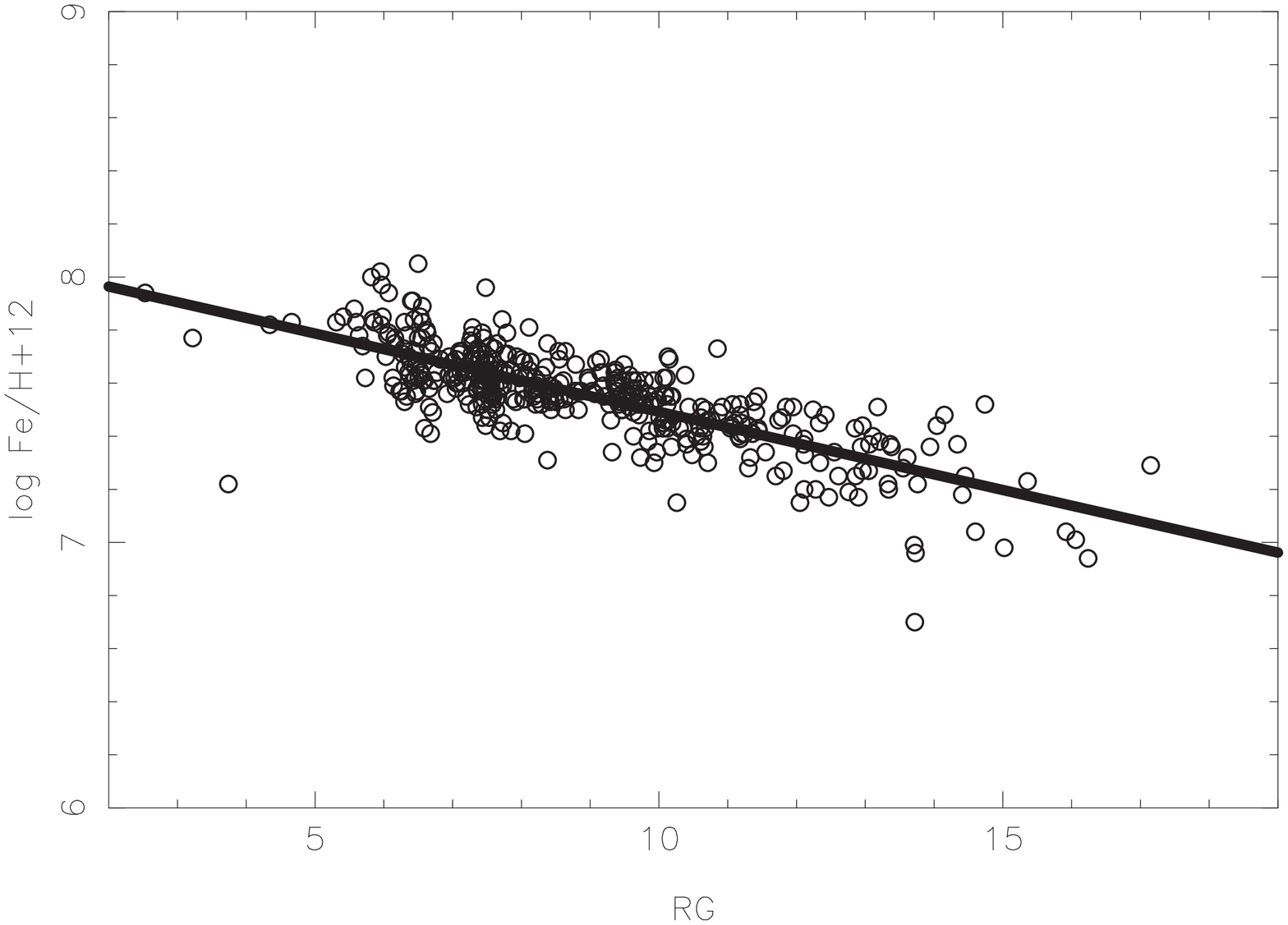} &
\includegraphics[height=7.0cm,angle=-90]{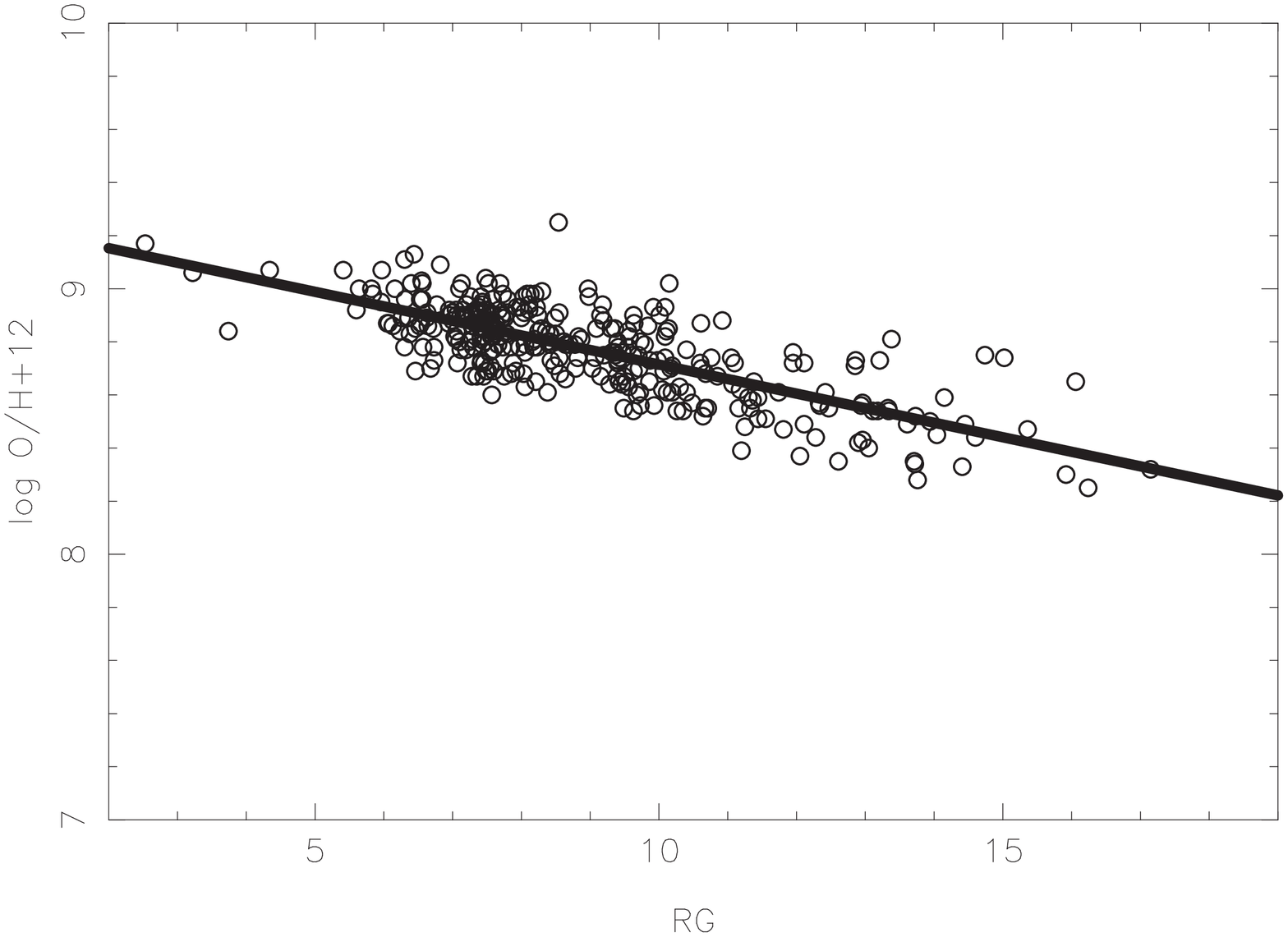} \\
\end{tabular}
\caption{Same as Figure 1 including the linear fits.}
\label{fig2ab}
\end{center}
\end{figure}

$$\log {\rm X/H} + 12 = A + B \ R_G \eqno(1)$$

\noindent
We notice that the O/H slope is very similar to the gradient found by \cite{korotin}, namely $-0.058$ 
dex/kpc, although we have used a larger sample. It can also be seen that both the Fe/H and O/H 
gradients are essentially the same within the uncertainties. From Figure~1ab there is some indication 
of a flattening in the inner Galaxy, where $R \leq 5\,$kpc, but the number of stars in this range is 
very small. Furthermore, the star with the lowest abundances at $R \simeq 3.6\,$kpc is probably a
W Vir star, so that this possibility should be viewed with caution.

As a second possibility, we have considered a double linear fit, in the ranges 4-11 kpc and 11-19 kpc, 
respectively, in agreement with the discussion by \cite{korotin}. The adoption of different ranges 
leads to slightly different results. The results of the fits are also given in Table~1 and Figures~3a 
and 3b. It can be seen that the correlation becomes less accurate for both segments, in view of the 
lower correlation coefficients $r$. For O/H there seems to exist some flattening in the outer galaxy, 
while for Fe/H the inner gradient is slightly flatter, but the difference is small, so that the 
gradient can be considered as essentially constant.

\begin{figure}
\begin{center}
\hspace{0.25cm}
\begin{tabular}{cc}
\includegraphics[height=7.0cm,angle=-90]{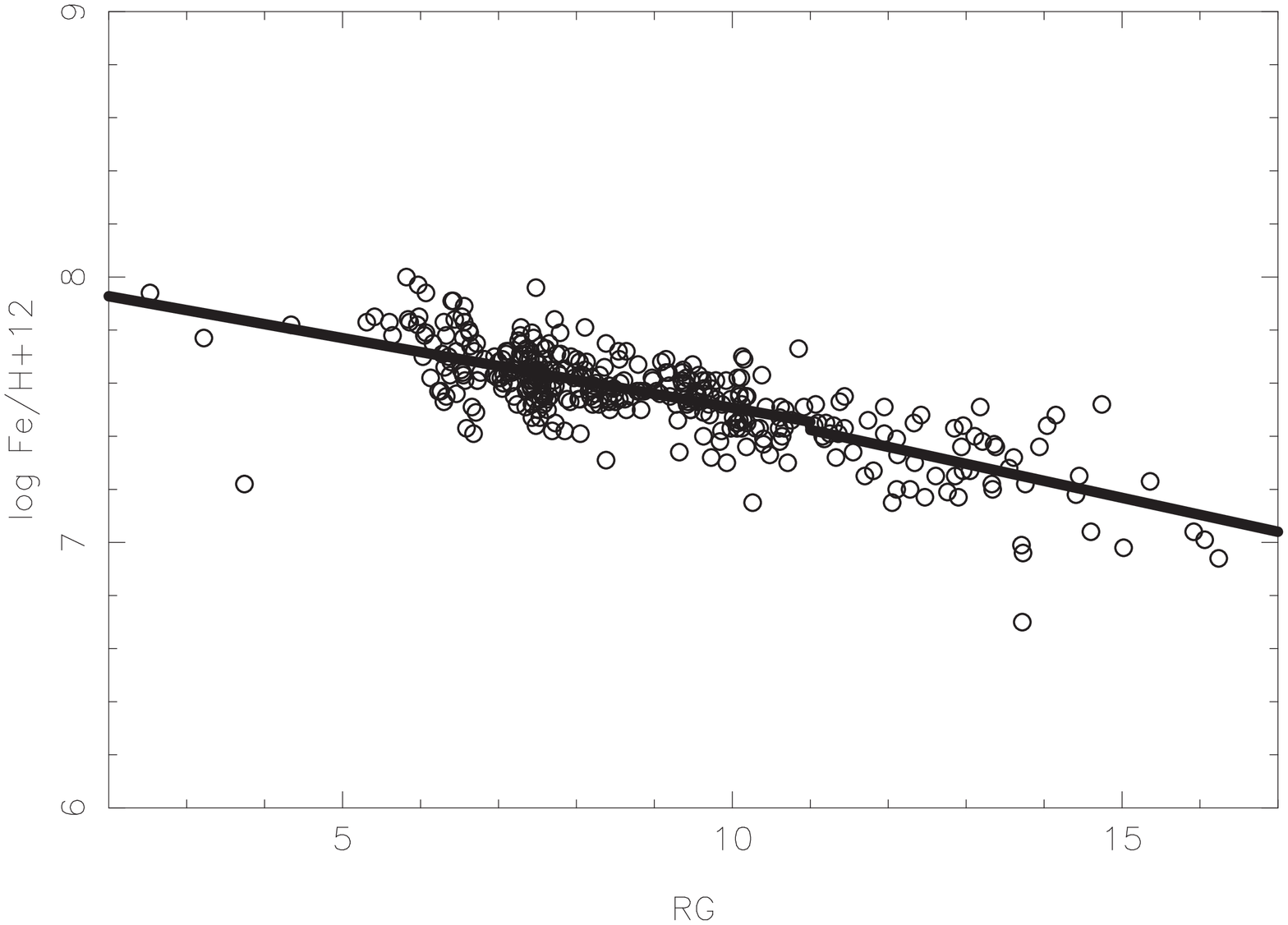} &
\includegraphics[height=7.0cm,angle=-90]{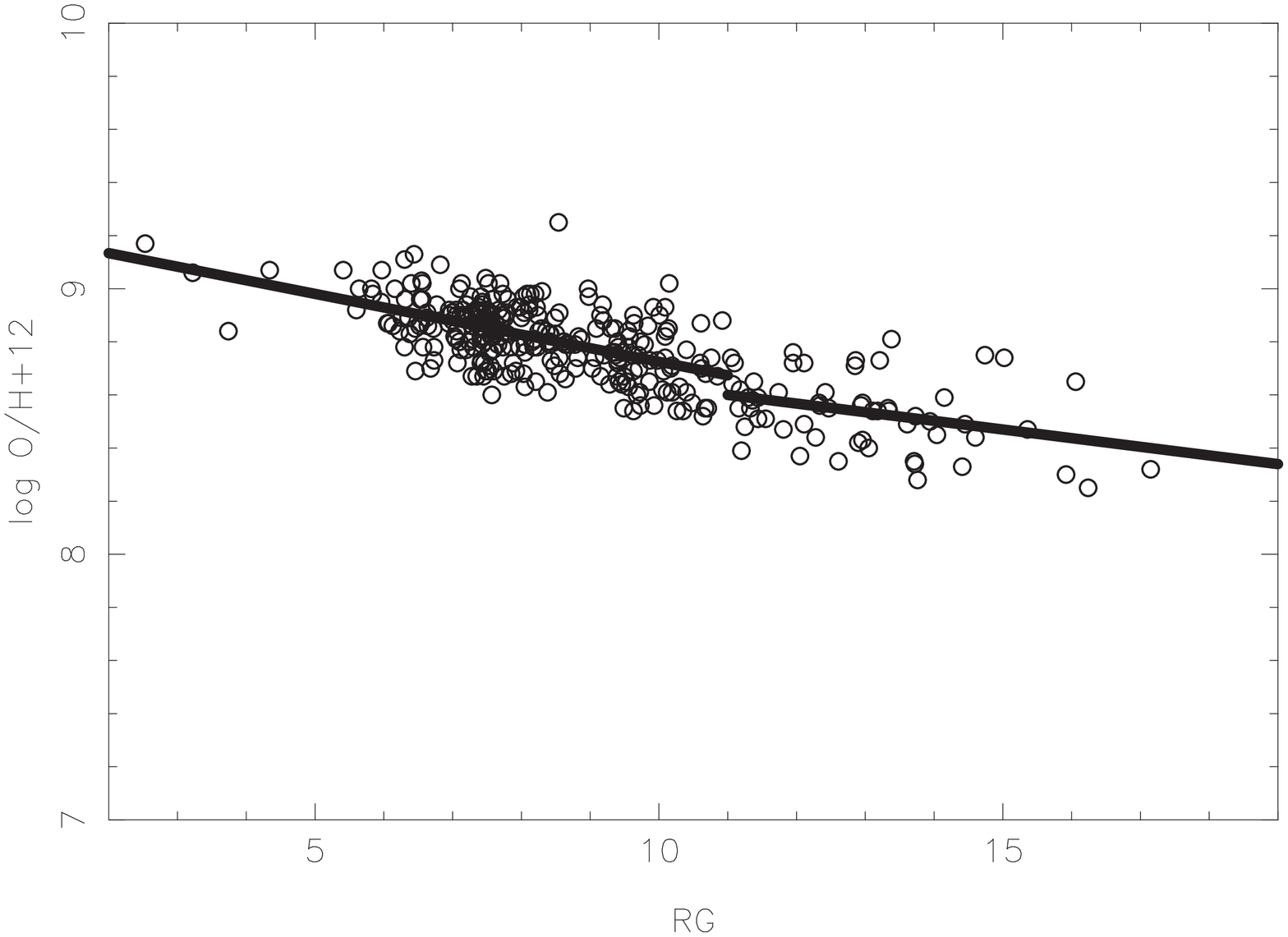} \\
\end{tabular}
\caption{Same as Figure 1 including double linear fits.}
\label{fig3ab}
\end{center}
%
\begin{center}
\hspace{0.25cm}
\begin{tabular}{cc}
\includegraphics[height=7.0cm,angle=-90]{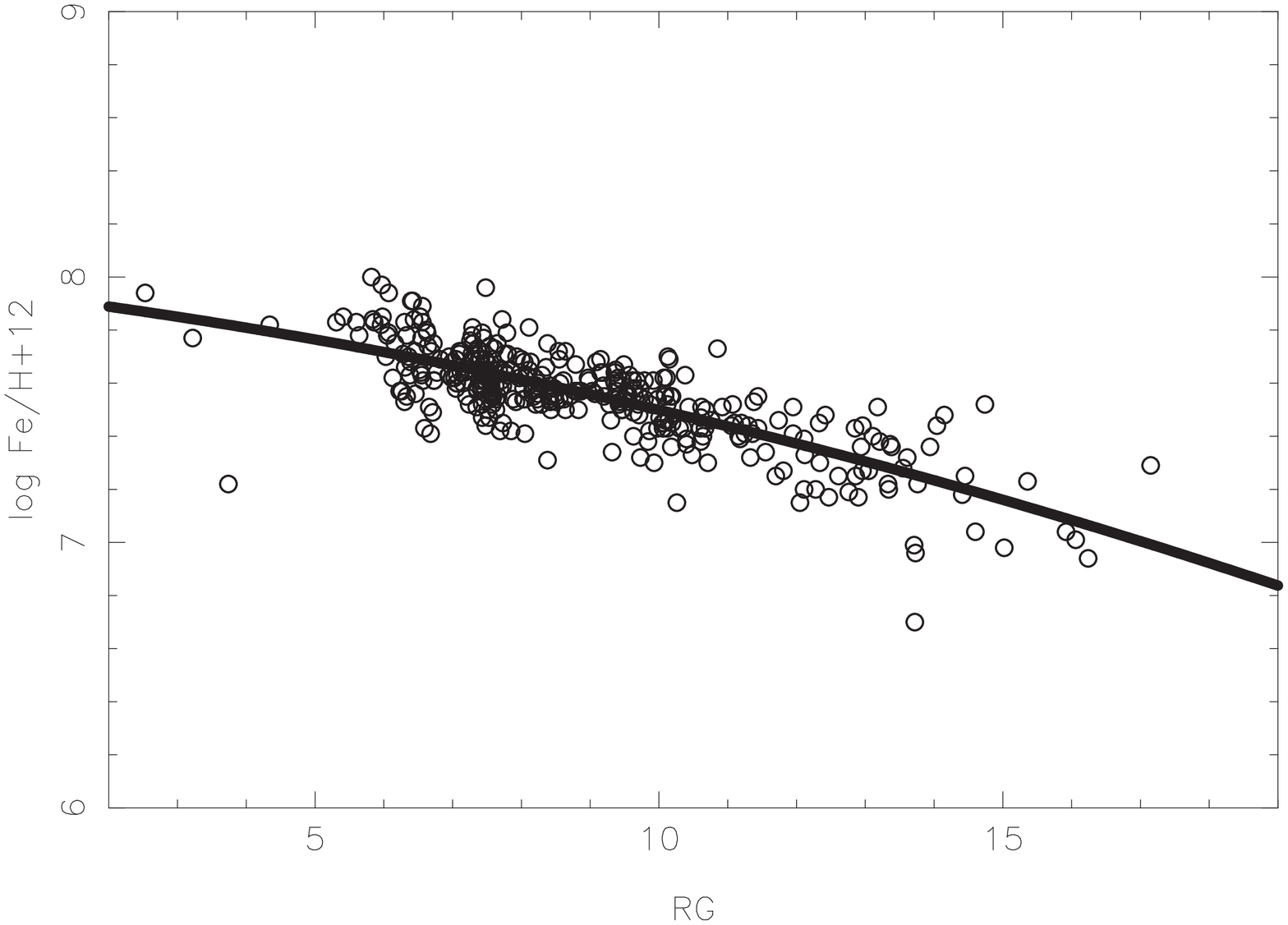} &
\includegraphics[height=7.0cm,angle=-90]{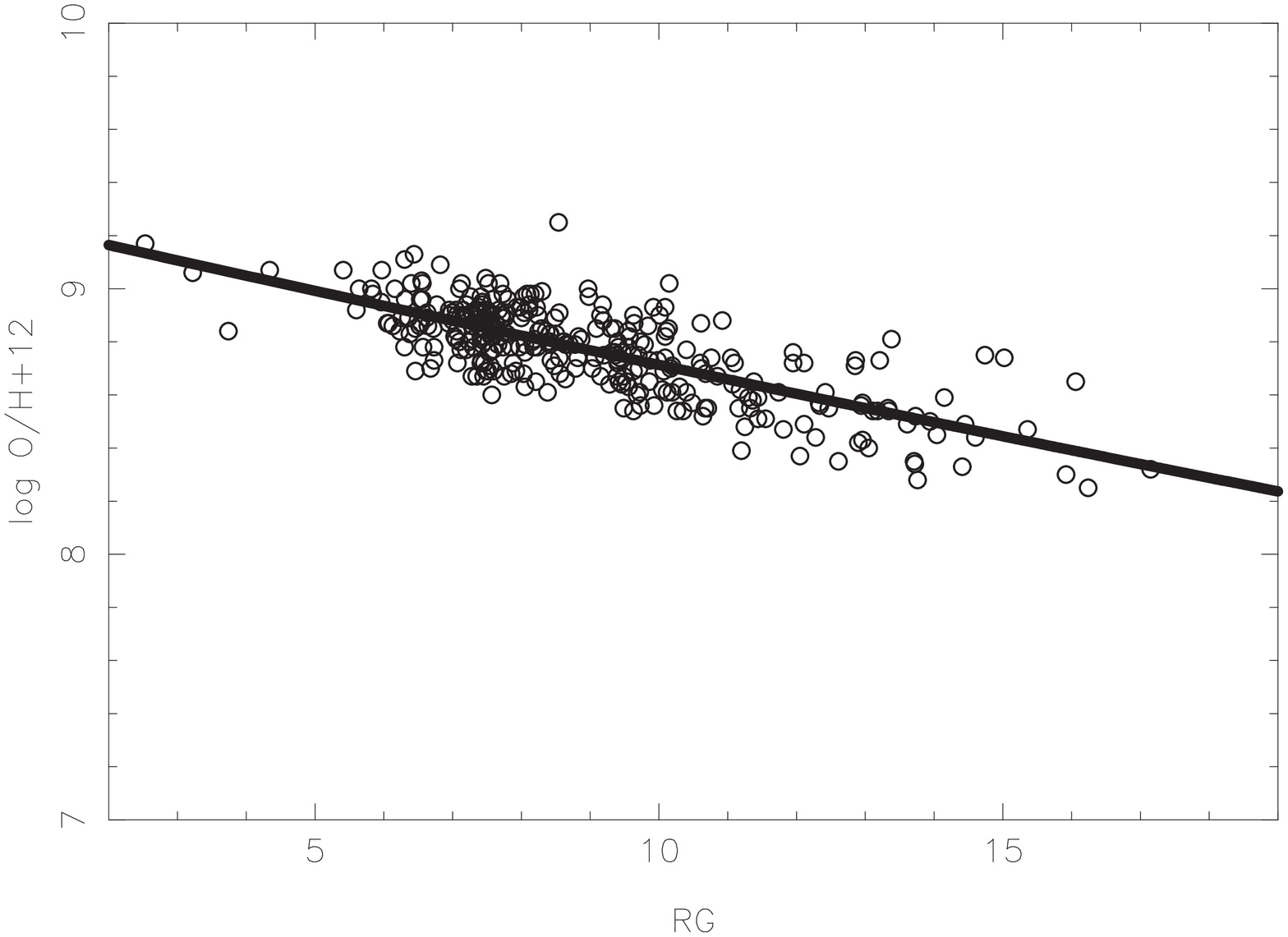} \\
\end{tabular}
\caption{Same as Figure 1 including quadratic polynomial fits.}
\label{fig4ab}
\end{center}
\end{figure}

As a third possibility, we have considered a quadratic polynomial fit, which in principle could take 
into account any detailed space variations of the gradients, especially in the outer galaxy.  The 
equation used is

$$\log {\rm X/H} + 12 = A + B \ R_G + C \ R_G^2 \eqno(2)$$

The results are shown in Figures~4ab, and are essentially  indistinguishable from the unique 
gradient shown in Figures~2ab. For example, at $R_G = 7.9\,$kpc we have $-0.056\,$dex/kpc and 
$-0.054\,$dex/kpc for the O/H and Fe/H gradients, respectively. The coefficients are given in  
Table~2. Again the correlations seem less accurate. For O/H the gradient is essentially constant, with 
maybe a slight flattening at large distances. For Fe/H there is a slight steepening in the outer 
galaxy, but again the difference is small. It should be mentioned that the results for the samples 
containing all variability phases and only one phase are similar, since the abundances in different 
phases are essentially the same. 

\bigskip

\centerline{{\baselineskip 12 pt {Table 1 - } 
Linear fits of O/H and Fe/H gradients for galactic cepheids.}}
$$\vbox{\halign{#\hfil&\quad\hfil#\hfil&\quad\hfil#\hfil
&\quad\hfil#\hfil&\quad\hfil#\hfil\cr\noalign{\hrule}
\ \cr
element & $n$ & $A$                  & $B$                & $r$          \cr
\ \cr
\noalign{\hrule}
\ \cr 
Single linear fits
\cr
O/H    & $331$ & $9.26 \pm 0.02$ & $-0.055\pm 0.003$ & $-0.75\pm 0.11$ \cr
Fe/H   & $361$ & $8.09 \pm 0.02$ & $-0.060\pm 0.003$ & $-0.77\pm 0.11$ \cr
\cr
Double linear fits
\cr
O/H: $R_G<11$ kpc  & $273$ & $9.24 \pm 0.04$ & $-0.051\pm 0.005$ & $-0.56\pm 0.10$ \cr
O/H: $R_G>11$ kpc  & $58 $ & $8.96 \pm 0.15$ & $-0.033\pm 0.011$ & $-0.35\pm 0.13$ \cr
                 &       &                 &                   &                 \cr
Fe/H: $R_G<11$ kpc & $298$ & $8.03 \pm 0.04$ & $-0.053\pm 0.004$ & $-0.58\pm 0.11$ \cr
Fe/H: $R_G>11$ kpc & $63$  & $8.13 \pm 0.17$ & $-0.064\pm 0.013$ & $-0.53\pm 0.15$ \cr
\ \cr
\noalign{\hrule}}}$$
\bigskip

\bigskip
\centerline{{\baselineskip 12 pt {Table 2 - } 
Quadratic fit of O/H and Fe/H gradients for galactic cepheids.}}
$$\vbox{\halign{#\hfil&\quad\hfil#\hfil&\quad\hfil#\hfil
&\quad\hfil#\hfil&\quad\hfil#\hfil&\quad\hfil#\hfil\cr\noalign{\hrule}
\ \cr
element & $n$ & $A$ & $B$  & $C$  & $\chi^2$   \cr
\ \cr
\noalign{\hrule}
\ \cr 
O/H  & $331$ & $9.281$ & $-0.05889$ & $-0.0002061$ & $0.0122$ \cr
Fe/H & $361$ & $7.957$ & $-0.03114$ & $-0.001462$  & $0.0130$ \cr
\ \cr
\noalign{\hrule}}}$$

\section{Gradients of photoionized nebulae}

Photoionized nebulae, namely HII regions and planetary nebulae are favourite objects to study radial 
gradients, especially for element ratios such as O/H, Ne/H, Ar/H and S/H, which are usually more 
difficult to determine in stars. On the other hand, Fe/H abundances are hardly measured in the 
nebulae, since most Fe is probably condensed in solid grains, so that the interpretation of the iron 
lines is more complex.

\subsection{HII regions}

Since the study by Shaver et al. (1983), many papers have dealt with the determination of radial 
gradients from HII regions, especially for O/H, on the basis of optical or infrared data [see for 
example \cite{esteban2018}, \cite{garcia}, \cite{esteban2017}, \cite{balser}, \cite{rudolph},  
\cite{quireza}, \cite{deharveng}]. These papers are based on rather small samples compared with the 
cepheid samples. Recent work suggests a constant gradient of about  $-0.04$ to $-0.05$ dex/kpc for 
O/H and no flattening at large galactocentric distances [\cite{esteban2018}]. The data are based on 
a sample with 35 objects with 10.4 GTC observations, for which electron temperatures have been 
measured. An inverse temperature gradient is also apparent. Some flattening has been pointed out in 
the inner Galaxy ($R_G \leq 5\,$kpc), a result also suggested on the basis of planetary nebulae 
[see for example \cite{gutenkunst}, \cite{ccm2011}]. This feature is also observed in our 
data, as mentioned in section~2.2. A compilation of recent determinations from photoionized nebulae 
can be found in \cite{molla2019}.

\subsection{Planetary Nebulae}

Planetary nebulae can be compared with HII regions, taking into account their different ages and also 
the fact that some elements have their abundances changed by the evolution of the progenitor stars, 
such as He and N. The determination of PN ages is a very difficult problem, which certainly affects 
the interpretation of the gradients [see for example \cite{stanghellini}, \cite{mrc2011, mci2010} for 
some recent discussions on the age problem]. 

Recent determinations of the abundance gradients include mainly the ratios O/H, Ne/H, S/H and Ar/H, 
with particular emphasis on the possible variations along the galactocentric distance and also 
relative to the distance from the galactic plane [see for example \cite{stanghellini}, 
\cite{pagomenos}, \cite{mcc2015, mc2013}]. A detailed study of distance-independent abundance 
correlations between Ne/H, S/H and Ar/H with O/H has been recently presented by \cite{mcc2017} for 
photoionized nebulae of the Local Group, showing that these elements are well correlated in HII 
regions and also in PN, albeit with a larger dispersion. The O/H radial gradients of PN are usually 
in the range $-0.02$ to $-0.05$ dex/kpc, and most recent papers favour the lowest gradients. However, 
the interpretation of these results is complex for a variety of reasons. First, the PN distances are 
not as well known as in the case of cepheids and even HII regions. Second, the (in)famous discrepancy 
between the results of forbidden lines and recombination lines is still largely unresolved, affecting 
both PN and HII regions [cf. \cite{carigi}]. Third, PN central stars have different ages, which are 
poorly known, especially for those objects older than about 4 Gyr. Fourth, radial migration has been 
recently found to be an important factor [see for example \cite{minchev}, \cite{jia}], which probably 
acts in order to flatten the gradients. This may explain some of the shallow gradients found in the 
literature, affecting the comparison between the gradients derived from young and old objects. 
Therefore the uncertainties of the average gradients are probably higher than generally assumed, and 
any spatial and temporal variations of the gradients based on PN should be viewed with care. 

\section{Some Conclusions}

$\bullet$ The radial gradients from cepheids are reasonably well represented by a unique gradient of 
about $-0.05$\ dex/kpc, which is essentially the same both for Fe/H and O/H. This is probably the best 
estimate of the radial gradient at the present time. A double linear fit does not seem likely, 
although some flattening in the outer galaxy can be observed for O/H. The data are consistent with
some flattening in the inner Galaxy, but this region is clearly not well covered in the present 
sample, since very few stars have galactocentric distances lower than about 5 kpc. 

\bigskip
$\bullet$ O/H  gradients from HII regions and cepheids are similar within the uncertainties. HII 
region abundances of O, Ne, S, and Ar show good correlations, and Ne, S, and Ar vary in lockstep with 
O, so that similar gradients can be expected for these elements.

\bigskip
$\bullet$ PN apparently have slightly flattened gradients of about $-0.03$\ dex/kpc for O/H and Ne/H, 
but most samples probably include objects with different ages.  The uncertainties in the age 
determination of the PN progenitor stars (and their distances) are very large for the older objects, 
with ages higher than about 5 Gyr. Radial migration makes things even more difficult, especially for 
older objects. These facts may explain the lower PN gradients compared with cepheids and HII regions.

\bigskip
$\bullet$
A comparison of the gradients derived from cepheids and photoionized nebulae shows as a first 
approximation that the gradients are similar, taking into account the uncertainties in the abundances, 
distances and, especially, the age determinations of PN. There are apparently no important temporal 
variations in the gradients in the last 3 to 5 Gyr for O/H. The distance-independent correlations for 
photoionized nebulae studied by \cite{mcc2017} and stars [see for example \cite{ramirez}] suggest 
that the Fe/H gradients are only marginally different from the O/H gradients.

\bigskip
$\bullet$
Recent theoretical models are able to explain the average gradients based on a inside out formation
scenario for the Galaxy. Some models can account for the inner flattening of the gradients, or predict
some steepening  at galactocentric distances larger than $R_0$ [see for example \cite{molla2019}, 
\cite{stanghellini}, \cite{grisoni}]. These models consider the time variation of the gradients and 
the results are often conflicting. The idea of a nearly constant gradient in the last few Gyr is 
supported by several models, but the behaviour of the gradients at earlier epochs is largely 
controversial. More recently, studies of the variation of the gradients with redshift are also 
consistent with approximately constant gradients for $z \leq 2$, but again the behaviour at earlier 
epochs is not clear, and a steeper gradient for $z > 3$ cannot be ruled out [see for example 
\cite{molla2019}, \cite{stanghellini}].

\acknowledgments This work was partially supported by CNPq (Process 302556/2015-0) and FAPESP 
(Process 2010/18835-3 and 2018/04562-7)

\end{document}